# Ethernet Networks: Current Trends and Tools

Abdulqader M. El-Sayed
Department of Computer Science
City University of New York
160 Convent Avenue, New York, NY 10031
Email: abdulqader.elsayed@cs.ccny.cuny.edu

*Abstract—*
**Ethernet topology discovery has gained increasing interest in the recent years. This trend is motivated mostly by increasing number of carrier Ethernet networks as well as the size of these networks, and consequently the increasing sales of these networks. To manage these networks efficiently, detailed and accurate knowledge of their topology is needed. Knowledge of a network's entities and the physical connections between them can be useful in various prospective. Administrators can use topology information for network planning and fault detecting. Topology information can also be used during protocol and routing algorithm development, for performance prediction and as a basis for accurate network simulations. From a network security perspective, threat detection, network monitoring, network access control and forensic investigations can benefit from accurate network topology information. In this paper, we analyze market trends and investigate current tools available for both research and commercial purposes.**

## I. Introduction

In recent years, Ethernet Networks have taken a new turn with respect to the marketplace. The sales of carrier Ethernet networks are on the rise, and collaborations with existing organizations are at an all-time high [20]. There have been considerable efforts in this industry to expand Metro Ethernet capabilities, which enable users to capture a greater amount of bandwidth at an affordable rate, while improving transparency for widespread use [20]. It is important for Ethernet developers and providers to capture the essence of this process through continuous improvements that are designed to increase bandwidth and other capabilities. In particular, carrier Ethernet functions are of critical importance in advancing growth, but this process requires extensive efforts from carrier operators in order to grow at the desired level [24]. There are important steps involved in promoting an effective approach to Ethernet growth, which increases the potential for expanded bandwidth and user friendly processes for all users.

Ethernet users should also look forward to advanced technologies, such as the 100Gbps card, which supports telecommunication industry providers for high-speed access and bandwidth [21]. This offering is only one of many emerging prospects for growth that are designed to lead the Ethernet industry way above its competition. A new approach to Ethernet networks capitalizes on the need for different types of visibility, while enabling network traffic to be controlled in a more efficient manner, thereby creating an environment that supports growth and traffic management [9].

## II. ETHERNET MARKET TREND

It is expected that by 2013, the market for Business Ethernet Services worldwide is expected to climb to $38.9 billion worldwide. As a result, there is an ever-increasing push for new Ethernet technologies that are designed to accommodate the needs of advanced users throughout the world. As key industry players continue to expand their efforts to adopt advanced Ethernet technologies, it is evident that this distinguished group will also seek simplicity and convenience in its Ethernet capabilities, such as the Vantage 8500, created by Voltaire Ltd. This strategy is important to many organizations not only to keep costs down, but to remain attractive to companies seeking new forms of revenue opportunity.

In spite of the advances being made in the Ethernet marketplace, some of its older offerings have suffered. For example, the 10 Gigabit Ethernet strategies has been weak; however, it is expected that this market will bounce back in response to higher demand in data centers that employ virtualization techniques [1]. These processes are of particular importance in advancing the industry to new heights, while also attempting to capitalize on new and inexperienced users that seek a quick fix to their bandwidth and speed problems.

Despite the economic recession, Ethernet capabilities and offerings continue to grow at steady rates. According to [12], "Carrier Ethernet is on a roll. Equipment sales are set to decline slightly due to increased competition and a weak US dollar, but Insight Research is projecting a 14 percent growth in service revenues this year, rising to 32 percent by 2012." In this context, it is important to consider how carrier Ethernet creates an environment that embraces change and high quality technologies, despite limited market growth. Because technology continues to attract new consumers and industries to the fold, it is possible for many Ethernet developers to capture some of the revenue and to increase their exposure through high-intensity capabilities and expanded bandwidth options.

Finally, new technologies such as the power-over-Ethernet (PoE) device [17] and extended warranties on Ethernet routing switches [16] represent new alternatives for the future of the Ethernet in order to expand capabilities and to capture and retain the market's attention in both industry and consumer circles, where demand is high and growth is steady. By expanding the research portfolio in this field, Ethernet growth is likely to continue to exceed expectations, and will promote new alternatives for companies and consumers seeking to increase their bandwidth and speed capabilities for both small and large-scale projects.

## III. DIFFERENT VIEWS OF ETHERNET NETWORK

Ethernet networks can be viewed in various prospective based on the authority's interest. In this section, we discuss different views of Ethernet networks.

### A. View of a business manager

Business managers are mostly interested in what devices are present in a network. It is not necessary for them to know how devices are connected. They have to know what devices are operating in a network and what devices will be needed in the future, so they can create budget plans of how much money must be available for buying new ones or exchanging obsolete ones. This point of view is based on a device inventory or stock.

Another possible demand of a business manager could be, to know where a customer's web-site can be seen, where this site is currently looked at or how often it is viewed. This is used for prediction and calculation of web-site statistics. In these cases it is also necessary to know the network topology. Otherwise wrong decisions are likely to be made based on incomplete or wrong knowledge. A less detailed topology view is needed in this case.

### B. View of a software engineer

For a software engineer of distributed applications it is elementary to have an imagination of the network topology, he designs his application for. Parameters that play a role are: throughput, latency and reachability for example. With this knowledge, possible bottlenecks or faulty operations can be avoided. Summarizing, this point of view relies on path-information and connections throughout a certain network. It is not important to know, what exact class of devices lay on the path passed through the network, but the constraints of this path are of interest.

### C. View of a network manager

The persons who are most interested in network topologies are network managers. It is their job to know the topology. They must jump in if any malfunction, caused by user, device or management errors occurs. In everyday situations they have to plan networks to be operating at their best possible efficiency. Reliability and fault tolerance are other important topics. As network requirements are constantly in change and to be prepared in any exceptional situation, it is highly eligible to know the current state of topology. At this point of view, both link parameters and connected devices are of interest.

## IV. EXISTING SOFTWARE IMPLEMENTING NETWORK DISCOVERY

This section discusses some products which can be placed in the area of network topology discovery. A division into open source and closed source product is made.

### A. Research Developed software

Layer-2 bridges are, also known as transparent bridges, can be invisible to other devices in the network. This property makes it hard to find layer-2 and respectively layer-1 network topologies. It can only be determined by using either special features, implemented to support network discovery or by exploiting and evaluating information needed for operation of bridges. Consequently, the topology information can be only obtained from resources such as STP-objects and forwarding data bases.

Notable amount of papers implemented these methods as described in this subsection. All of these tools are vendor-independent and based on public standards.

Both papers "Topology Discovery for Large Ethernet Network" [19] and "Topology Discovery in Heterogeneous IP Networks" [6, 7] rely on MAC-address learning of layer-2 bridges. They collect all entries from all known devices and apply certain rules on the derived sets of forwarding databases. Results from both methods are layer-2 network topologies, which consist of all devices having a distinct MAC-address. This includes end stations, too. The latter technique introduces the "Completeness Requirement". That means, every address forwarding table in each device must be complete. That is, they have to contain the full set of MAC addresses which are potentially reachable from any devices interface within a single subnet. As the standard aging time of address forwarding entries, kept by layer-2 bridges, is 300s, this constraint is hard to satisfy. But two proposals are made to accomplish the completeness.

First a constant network traffic is generated, which prevents forwarding entries from being aged out. This is proposed to be done by constantly sending Internet Control Message Protocol (ICMP) echo-request throughout the entire network and expecting the devices to return ICMP echo-replies. To gain permanent access to any machine in a large network and having them running the ICMP traffic generation all-time, is a challenging task itself. Second solution to soften the completeness requirement is to decrease the forwarding set of a bridge-port to a user-defined reasonably large fraction.

The second method of network topology discovery based upon address forwarding tables uses an opposite constraint. It defines the "Simple Connection Theorem", which implies a minimum knowledge constraint. For a pair of bridges, only three forwarding entries have to be shared and only one host has to be accessed, namely the one discovery queries are sent and received from. All this reduces additional effort for discovering network topology.

A third method for discovery based on bridges functioning is the "Layer-2 Path Discovery Using Spanning Tree MIBs" [27]. As the name suggests, this discovery uses the STP-data stored in each layer-2 capable device to ascertain network topology.

Each bridge by default transmits Bridge Protocol Data Units (BPDUs) which contain Spanning Tree Protocol information from the sending bridge, including among other values the Bridge Identifier (BID) and port identifier for the sending bridge. The bridge-ID consists of eight eight-bit values. Two octets define priority followed by six octets, which are recommended to be equal to the lowest numbered bridge-port MAC-address, i.e. the address for port 1. The transmission of STP configuration frames are repeated periodically and after a certain time the spanning-tree converges to a stable state.

Each bridge stores the STP-data in its SNMP-Bridge-MIB. By querying all bridges it is possible to obtain the layer-2 network topology. An advantage of this method is that even ports in STP-blocking state can be detected, as the information about the port is also advertised. A problem occurred during test, Cisco Catalyst 5000 switches did not store the port-ID value, so with this class of devices it is not possible to apply the STP discovery method.

In the paper "Physical Topology Discovery for Large Multi-Subnet Networks", Bejerano *et al.* [3] investigated topology discovery for Ethernet networks in presence of uncooperative elements such as hubs. Bejerano described an algorithm to discover the topology in such case. However, neither complexity analysis nor implementation of their algorithm was described.

Another work by Bejerano [4] "Taking the Skeletons Out of the Closets: A Simple and Efficient Topology Discovery Scheme for Large Multisubnet Networks" showed the limitations of the algorithms developed by Lowekamp *et al* [19] and Breitbart *et al* [6, 7] in multi-subnet networks or in the presence of uncooperative devices. The algorithm proposed in this paper was implemented and discovered the topology in most cases.

In the previous papers, no assumption was made about uniqueness of discovered topology or how to handle a potential situation when the forwarding database can represent more than one topology. I their papers "Characterization of layer-2 unique topologies" and "Characterization of Layer-2 Unique Topologies in Multisubnet Local Networks", Breitbart and Gobjuka [8, 13] investigated that issue and proposed an algorithm to determine whether the forwarding database can represent one or more Ethernet topologies. However, this algorithm could function only if all devices have accessible MIB and all forwarding data base of all ports is complete.

In a follow-up research, Gobjuka *et al.* [15] extensively studied in their paper "Discovering Network Topology of Large Multisubnet Ethernet Networks" the situation where some devices in the Ethernet network prohibit accessing to their MIBs. They proposed what they called Reachability Set as replacement of forwarding databases and they extended forwarding databases to obtain complete set of reachable devices from each device port. Then, they proposed algorithm to discover the Ethernet network topology and determine the uniqueness. Finally, they showed that their algorithm can map all possible topologies.

An extensive analysis for the case when forwarding databases of device ports are incomplete was published in the paper "Ethernet Topology Discovery for Networks with Incomplete Information" by Gobjuka *et al.* [14]. The authors provided detailed analysis of previous work in literature and demonstrated cases where previously published methods cannot return any network topology. They also explained several rules called "extension rules" to complete forwarding databases of network device ports by applying graph theory concepts. They showed that discovering the tomography of a single subnet network can be intractable by reducing from the betweenness problem when the "separation rule" is valid. They also showed how the difficulty of obtaining the topology turns to be polynomial-time solvable when the separation rule is not satisfied. And that trying to determine whether more than one tomography can be obtained from the forwarding database is not polynomially solvable even if a network topology is provided. The validity of their algorithms was shown by implementations and simulation results.

All previous vendor-independent methods mentioned, assume that all necessary data contained in devices, are read out before the network topology can be recreated. This arises from the fact that ports and devices are identified for topology determination by their layer-2 addresses respectively their bridge-ID. Network-management tools like SNMP or telnet use Internet Protocol (IP) addresses or Domain Name System (DNS) names to address devices and they communicate using higher layer (3, 4 and 7) protocols.

"Find devices by MAC-addresses", it is based on the idea that port-MAC-addresses are consecutive. Using this assumption it is possible to identify a device a port belongs to with a certain probability.

A further issue arises by the usage of VLANs. As the name implies, VLANs are virtual LANs, which determines that each VLAN can have its own set of forwarding entries and Spanning Tree data. [IEEE802.1Q] 2003 edition introduces multiple Spanning Tree instances.

Beside all the method-specific cutbacks, it must be clear that it is only possible to discover an "active" topology. Which means only links that are enabled and are transporting Ethernet-frames can be found by any topology detection method. A port in state "administratively down" is comparable to a physically not connected port, so it cannot transmit any bit of information. Consequence of that is, a pair of ports connected to each other and one port is disabled, will not be discovered, although it is part of the layer-1 topology.

### B. Industry Developed Commercial Software

Industry has had also its own share in developing network management tools. However, most of these tools were more of node managers than topology discovery tools. Examples for industry developed commercial tools are mentioned in this subsection. Cisco's CiscoWorks – Campus Manager [10] and Hewlett-Packard's OpenView - Network Node Manager [28] are among the most common in market. Both offer a vast amount of management functions beside the network discovery and have options to enhance to features desired. However, the topology discovery functionalities provided in both tools are very simple and fail often to return the accurate topology. A detailed description would be far beyond the scope of this work. Both products probably can handle all the necessary situations with respect to device management, but in a university environment there are highly qualified employees available, who are able to maintain and handle self-developed solutions and students who enhance existing products.

## V. CONCLUSION

Accurate knowledge of network topology can have a strong impact on several management and performance related issues. Investigation of the topological characteristics of Ethernet networks and its practical uses is gaining increasing attention due to the steady Ethernet market horizontal and vertical growth. In this paper, in depth analysis of market trends and tools available for Ethernet network management is presented.

Ethernet network discovery tools and methods developed by research community are diverse and can do tasks from purely topology discovery to network mapping, and they can function in single subnet or multisubnet networks that may or may not provide access to their MIBs. Some of these tools can also discover the topology when obtained information is incomplete. From industry prospective, most of the tools provide complete node management functionality including node discovery, service monitoring, port scanning and even vulnerability scanning but they have very simple network topology discovery capabilities.

Discovering physical IP network connectivity is not easy task and despite the critical role of topology information in enhancing the manageability of modern IP networks, none of the network management platforms currently available on the market can offer a general-purpose tool for automatic discovery of physical IP network connectivity.